\documentstyle[preprint,aps]{revtex}
\begin{document}
\preprint{LA-UR-92-3726-REV}
\draft
\widetext
\title{
A Bargmann-Wightman-Wigner Type Quantum Field Theory
\footnotemark[1]
\footnotetext[1]{This work was done under the auspices of the U. S.
Department of Energy.}
}
\author{D. V. Ahluwalia \dag, M. B. Johnson \dag and \ T. Goldman \ddag}
\address{\dag MP-9, MS H-846, Nuclear and Particle Physics Research Group\\
 Los Alamos National Laboratory,
Los Alamos, New Mexico 87545, USA}

\address{\ddag T-5, MS B-283, Medium Energy Physics Theory Group\\
 Los Alamos National Laboratory,
Los Alamos, New Mexico 87545, USA}
\maketitle

\begin{abstract}
We show that the $(j,0)\oplus(0,j)$ representation space associated with
massive particles is a concrete realisation of a quantum field theory,
envisaged many years ago by Bargmann, Wightman and Wigner, in which bosons and
antibosons have opposite relative intrinsic parities. Demonstration of the
result requires a careful {\it ab initio} study of the $(j,0)\oplus(0,j)$
representation space for massive particles, introducing a wave equation with
well defined transformation properties under $C$, $P$ and T, and addressing the
issue of nonlocality required of such a theory by the work of Lee and Wick.
\end{abstract}

\newpage

While most of the specific theoretical questions of hadronic structure and
interactions must be decided within the framework of quantum chromodynamics,
there remain certain aspects which depend only on the constraints imposed by
Poincar\'e covariance. Many years ago, Wigner \cite{EW}  provided the basic
framework for the Poincar\'e covariant considerations. The essential elements
of these considerations are the kinematical symmetries (continuous Poincar\'e
symmetries and space time reflections) and the behaviour of quantum mechanical
states under these transformations. From these follow certain general
characteristics such as equal masses and relative intrinsic parities of
particle and antiparticle pairs. Such an approach may therefore have utility
in establishing the general framework of an effective field theory of hadrons,
and such considerations have, in fact, motivated the pioneering work of
Weinberg \cite{SW} on field theories in a specific Lorentz group
representation, $(j,0)\oplus(0,j)$, of spin-j particles, as well as recent
extensions of this work \cite{DVA}.

Although there has been considerable work on the $(j,0)\oplus(0,j)$
representation with specific applications in mind, additional insight into this
representation can be obtained by considering it as a special case of the
general classification of quantum field theories by Wigner [1b], in which he
distinguishes four classes. The class of theories in which  a boson and
anti-boson have opposite intrinsic parities are informally known as
``Wigner-type,'' but in view of Wigner's note [1b, p. 38]
 that ``much of the'' work in Ref.
[1b] ``was taken from a rather old but unpublished manuscript by V. Bargmann,
A. S. Wightman and myself,'' we take the liberty of calling this type of theory
 Bargmann-Wightman-Wigner type (BWW-type) quantum field theory. Even though the
generality of BWW's arguments  is remarkable, at present no explicit
construction of a BWW-type quantum field theory is known to exist. Nor has it
been realised that the $(j,0)\oplus(0,j)$ representation for massive particles
is a realisation of the BWW-type quantum field theory. In this paper we show
that this is the case. We do this by considering the special case of the
$(1,0)\oplus(0,1)$ field and working out explicitly its properties under C, P
and T.

We begin with a brief review of $(j,0)\oplus(0,j)$ representation space.
In the notation of  Refs. \cite{DVA,LR} $(j,0)$ and  $(0,j)$
spinors have the following Lorentz transformation properties
\begin{eqnarray}
&&(j,0):\quad \phi_R(\vec p\,)\
=\,\Lambda_R\,\phi_R(\vec 0\,) \,=\,\exp\left(+\,\vec
J\cdot\vec\varphi\,\right)\,\phi_R(\vec 0\,)\quad,\label{r}\\
&&(0,j):\quad\phi_L(\vec p\,)\,
=\,\Lambda_L\,\phi_L(\vec 0\,) \,=\,\exp\left(-\,\vec J\cdot\vec\varphi\,
\right)\,\phi_L(\vec 0\,)\quad. \label{l}
\end{eqnarray}
 The $\vec J$ are the standard $(2j+1)\times(2j+1)$
angular momentum matrices, and $\vec \varphi$ is the boost parameter
defined as
\begin{equation}
\cosh(\varphi\,)
\,=\,\gamma\,=\,{1\over\sqrt{1-v^2}}\,=\,{E\over m},\quad\quad \sinh(
\varphi\,)\,=\,v\gamma\,=\,{| \vec p\, |\over m},\quad\quad\hat\varphi={\vec  p
\over| \vec p\,|},
\end{equation}
with $\vec p$ the three-momentum of the particle. In order to stay as close as
possible to the standard treatments \cite{LR}  of the $(1/2,0)\oplus(0,1/2)$
 Dirac field, we now
introduce a generalised canonical representation [3a,3f]
$(j,0)\oplus(0,j)$-spinor
\begin{equation}
\psi(\vec p\,)\,=\,{1\over\sqrt{2}}
\left(
\begin{array}{cc}
\phi_R(\vec p\,)\,+\,\phi_L(\vec p\,)\\
\phi_R(\vec p\,)\,-\,\phi_L(\vec p\,)\\
\end{array}\right)\quad,\label{crs}
\end{equation}
In the  $(j,0)\oplus(0,j)$ representation space there are $(2j+1)$
``$u_\sigma(\vec p\,)$ spinors'' and $(2j+1)$ ``$v_\sigma(\vec p\,)$ spinors.''
As a consequence of the  transformation properties (1,2),
and the definition (\ref{crs}), these spinors transform as
\begin{equation}
\psi(\vec p\,)\,=\,M(j,\,\vec p\,)\,\psi(\vec 0\,)\,=\,
\left(
\begin{array}{ccc}
\cosh(\vec J\cdot\vec\varphi)&{\,\,} &\sinh(\vec J\cdot\vec\varphi) \\
\sinh(\vec J\cdot\vec\varphi)&{\,\,} &\cosh(\vec J\cdot\vec\varphi)
\end{array}\right)\,\psi(\vec 0\,)\quad.
\label{b}
\end{equation}
If we work in a representation of the $\vec J$ matrices in which $J_z$ is
diagonal, then the rest spinors
$u_\sigma(\vec 0\,)$ and $v_\sigma(\vec 0\,)$,
$\sigma=j,j-1,\cdots,-j$, can be written in the form of
$2(2j+1)$ single column matrices each with $2(2j+1)$ elements as follows
\begin{equation}
u_{+j}(\vec 0\,)\,=\,
\left(
\begin{array}{c}
N(j)\\
0\\
\vdots\\
0
\end{array}\right)
\,,
u_{j-1}(\vec 0\,)\,=\,
\left(
\begin{array}{c}
0\\
N(j)\\
\vdots\\
0
\end{array}\right)
\quad,
\cdots,\quad
v_{-j}(\vec 0\,)\,=\,
\left(
\begin{array}{c}
0\\
\vdots\\
0\\
N(j)
\end{array}\right)\quad.\label{rs}
\end{equation}
For convenience, and so that the rest spinors vanish in the $m\rightarrow 0$
limit, we choose the normalisation factor $N(j)=m^j$.
With this choice of the normalisation
the spinors $u_\sigma(\vec p\,)$ and
$v_\sigma(\vec p\,)$ are normalised as follows:
$\overline{u}_\sigma(\vec p\,)\,u_{\sigma'}(\vec
p\,)=m^{2j}\delta_{\sigma\sigma'}$ and
$\overline{v}_\sigma(\vec p\,)\,v_{\sigma'}(\vec
p\,)=-m^{2j}\delta_{\sigma\sigma'}$, with
$
\overline{\psi}_\sigma(\vec p\,)\,=\,
{\psi}_\sigma^\dagger(\vec p\,)\,\Gamma^0\,.
$
 Here $\Gamma^0$ is a block diagonal matrix with $(2j+1)\times(2j+1)$
identity matrix $I$ on the upper left corner and $-I$ on the  bottom right
corner.
The reader would
immediately realise that for $\vec J=\vec\sigma/2$, $\vec\sigma$ being Pauli
matrices,  the boost $M(1/2,\,\vec p\,)$ is identical with the canonical
representation boost \cite{LR}  for the $(1/2,0)\oplus (0,1/2)$
Dirac spinors. While investigating the $C$, $P$ and
T properties we will consider the
$(1,0)\oplus(0,1)$ field as an example case. Hence, we present the explicit
expressions for the $(1,0)\oplus(0,1)$ spinors
(in the generalised canonical
representation introduced above). The three $u_{0,\pm 1}(\vec p\,)$
and the three $v_{0,\pm 1}(\vec p\,)$ spinors read:
{\footnotesize
\begin{eqnarray}
u_{_ {+1}}( \vec p\,)&\,=\,&
\pmatrix{m+\left[(2p_z^2\,+\,p_{_{+}} p_{_{-}}) / 2(E+m)\right]\cr
                      {p_z p_{_{+}}/{\sqrt 2}(E+m)}\cr
              { p_{_{+}}^2/ 2(E+m) }\cr
               p_z\cr
                   {p_{_{+}}/{\sqrt 2}}\cr
                   0\cr}\,,\quad
u_{_{0}}( \vec p\,)\,=\,\pmatrix{{p_z p_{_{-}}/{\sqrt 2}(E+m)}\cr
                      m+\left[{p_{_{+}} p_{_{-}}/(E+m) }\right]\cr
                       -{p_z p_{_{+}}/{\sqrt 2}(E+m)}\cr
                       {p_{_{-}}/{\sqrt 2}}\cr
                          0\cr
                        {p_{_{+}}/{\sqrt 2}}\cr}\quad,\nonumber\\
u_{_{-1}}( \vec p\,)&\,=\,&\pmatrix{ { p_{_{-}}^2/ 2(E+m) }\cr
                             -{p_z p_{_{-}}/{\sqrt 2}(E+m)}\cr
                   m+\left[{(2p_z^2\,+\,p_{_{+}} p_{_{-}})/ 2(E+m)}\right]\cr
                      0\cr
                      {p_{_{-}}/{\sqrt 2}}\cr
                   -p_z\cr}\,,\quad\qquad
v_\sigma(\vec p\,) \,=\,
\left(
\begin{array}{ccc}
0 & {\quad}I \\
I & {\quad}0
\end{array}\right)
\,u_\sigma(\vec p\,)\label{uv}\quad.
\end{eqnarray}
}
In the above equation we have defined $p_\pm\,=\,p_x\,\pm\,i\,p_y$.

Once we obtain the $(j,0)\oplus(0,j)$ spinors we make use Weinberg's \cite{SW}
observation that the general form of a field operator is dictated upon us by
arguments of Poincar\'e covariance without any explicit  reference to a wave
equation. The $(j,0)\oplus(0,j)$ field operator reads:
\begin{equation}
\Psi(x) \,= \,
\sum_{\sigma=+j}^{-j}
\int {d^3p\over (2\pi)^{3} } {1\over 2\,\omega_{\vec p}}
\Big[ u_\sigma(\vec p\,)\, a_\sigma(\vec p\, )\, e^{-i p\cdot x}
+  v_\sigma(\vec p\,) \,b^\dagger_\sigma(\vec p\,) \,e^{i p\cdot x} \Bigr ]
\quad,\label{fo}
\end{equation}
where $\omega_{\vec p\,} = \sqrt{m^2 + {\vec p\,}^2}$; and
$[a_\sigma(\vec p\,),\,a^\dagger_{\sigma'}(\vec p^{\,\prime}]_\pm=
\,\delta_{\sigma\sigma'}
\delta(\vec p-\vec p^{\,\prime})$.

Since so far no wave equation  has been invoked it is important to see how one
may obtain a wave equation.
To derive the wave equation satisfied by the
$(j,0)\oplus(0,j)$ spinors we
observe that the general structure of the rest spinors given by Eq.
(\ref{rs}) implies that :
$
\phi_R(\vec 0\,)\, \,=\,\wp_{u,v}\,\phi_L(\vec 0\,)\,,
$
with
$
\wp_{u,v}= +1$ for the $u$-spinors and
$
\wp_{u,v}= -1$ for the $v$-spinors.
It may be parenthetically noted that in a similar context for spin one half,
Ryder [4, p.44] assumes the validity of an equation which reads,
$
\phi_R(\vec 0\,)=\phi_L(\vec 0\,)\,,
$
on the grounds that
``when a particle is at rest, one cannot define its spin as either left- or
right-handed.'' However, as we note this is simply a consequence of the general
structure of our theory --- moreover, in the process we find an additional
minus sign (in the $\wp_{u,v}$ factor). This factor would be found to have
profound significance for the internal consistency and consequences  of our
study.
 When
we couple the relations
$
\phi_R(\vec 0\,) \,=\, \wp_{u,v}\,\phi_L(\vec 0\,)\,
$
with the transformations properties
(1) and (2) we [3f] obtain
$
\left[
\gamma_{\mu\nu\ldots\lambda}\,p^\mu p^\nu\ldots p^\lambda
\,-\,{\wp_{u,v}}\,m^{2j} I\right]\,\psi(\vec p\,)\,=\,0\,.
$
This equation, except for the  factor of ${\wp_{u,v}}$ attached to the
mass term, is identical to the Weinberg equation \cite{SW} for the
$(j,0)\oplus(0,j)$ spinors. The $2(2j+1)\times 2(2j+1)$ dimensional
$\gamma_{\mu\nu\ldots\lambda}$ matrices, with $2j$ Lorentz indices,
which appear here can be found in Ref.
\cite{SW}, or in more explicit form in Ref. [3f].
For $j=1/2$ case, this wave equation
 is found to be identical to the Dirac equation
in momentum space. For the $(1,0)\oplus(0,1)$
configuration-space-free-wave-functions
$\psi(x)\equiv \psi(\vec p\,) \exp(-i\wp_{u,v}\,p\cdot x)$,
this wave equation  becomes
\footnotemark[1]
\footnotetext[1]{Note that the $(1/2,0)\oplus(0,1/2)$ wave functions satisfy
$(i\gamma_\mu\partial^\mu\,-\,m\,I)\,\psi(x)\,=\,0$, {\it without} the
${\wp_{u,v}}$ attached to the mass term. It is readily seen that this
``cancellation'' of the ${\wp_{u,v}}$ when going from momentum space
$\rightarrow$ configuration space occurs for fermions, but not for bosons,
in general.}
\begin{equation}
\left(
\gamma_{\mu\nu}\partial^\mu\partial^\nu\,+\,
\wp_{u,v}\,m^2
\right)
\,\psi(x) \,=\,0\quad,\label{ea}
\end{equation}
with generalised canonical representation expressions for the $6\times 6$
ten
$\gamma_{\mu\nu}$-matrices given by
\begin{eqnarray}
\gamma_{00}\,=\,
\left(
\begin{array}{ccc}
I &{\,\,} &0\\
0 &{\,\,} &-I
\end{array}
\right),\quad
\gamma_{\ell0}\,=\,\gamma_{0\ell}\,=\,
\left(
\begin{array}{ccc}
0 &{\,\,}& -J_\ell\\
J_\ell&{\,\,} & 0
\end{array}
\right)\quad, \nonumber \\ \nonumber \\
\gamma_{\ell\jmath}\,=\,\gamma_{\jmath\ell}\,=\,
\left(
\begin{array}{ccc}
I & {\,\,}&0\\
0 &{\,\,}&-I
\end{array}
\right)
\,\eta_{\ell\jmath} \,+\,
\left(
\begin{array}{cc}
\{J_\ell,J_\jmath\} & 0\\
0 & -\,\{J_\ell,J_\jmath\}
\end{array}
\right)\quad.
\label{gm}
\end{eqnarray}
Here $\eta_{\mu\nu}$ is the
flat space time metric with $diag\,(1,\,-1,\,-1\,-1)$; and
$\ell,\,\jmath$
run over the spacial indices $1, \,2, \,3$.

Equation (\ref{ea}) has three $u_\sigma(\vec p\,)$ and
three $v_\sigma(\vec p\,)$ solutions given by Eq. (\ref{uv}). However
these solutions can also be interpreted as associated with not only
Einsteinian
$E=\pm\sqrt{{\vec p}^{\,2}+m^2}$ but also with tachyonic [3e,3f]
$E=\pm\sqrt{{\vec p}^{\,2}-m^2}$. This can be readily inferred
[3e,3f]
by studying the $12th$ order polynomial in $E$:
$Det \,\left(
\gamma_{\mu\nu}p^\mu p^\nu\,+\,
\wp_{u,v}\,m^2\,I
\right)=0\,$.
The tachyonic solutions are at this stage unphysical and can be ignored as long
as interactions are introduced in such a manner that they  do not induce
transitions between the  physical and unphysical solutions. In this context,
for the spin one spinors we introduce
\begin{equation}
P_u\,=\,{1\over {m^2}}\sum_{\sigma=0,\pm 1} u_\sigma(\vec p\,)\,
\overline {u}_\sigma(\vec p\,)\,,\quad
P_v\,=\,-\,{1\over {m^2}}\sum_{\sigma=0,\pm 1} v_\sigma(\vec p\,)\,
\overline {v}_\sigma(\vec p\,)\quad,\label{pupv}
\end{equation}
and verify that $P_u^2=P_u$, $P_v^2=P_v$ and $P_u\,P_v=0\,$.

In order to establish that the field operator defined by Eq. (\ref{fo})
describes a quantum field theory of the BWW-type we now show
that bosons and antibosons, within the $(j,0)\oplus(0,j)$ framework
developed above,  indeed have opposite intrinsic parity  and well
defined $C$ and $T$ characteristics.
We begin with the classical considerations similar to the
ones found for the $(1/2,0)\oplus(0,1/2)$ Dirac field in the standard
texts, such as Ref. \cite{ON}. As the simplest example case we study the
$(1,0)\oplus(0,1)$ field in detail. As such we seek a
parity-transformed wave function
$
\psi'(t',\,{\vec x}\,') \,=\,S(P)\,\psi(t,\,\vec x)\,;\quad
{x'}^\mu \,=\,{\left[\Lambda_P\right]^\mu}_\nu\,x^\nu\,,
$
 [Here: $\Lambda_P=diag\,(1,\,-1,\,-1,\,-1)$ so that $t'=t$ and
$\vec x^{\,\prime}=-\vec x$]
such that Eq. (\ref{ea}) holds true for $\psi^\prime(t',\,{\vec x}\,')$.
That is:
$
\left(\gamma_{\mu\nu}\,\partial^{\,\prime\mu}\,\partial^{\,\prime\nu}
 \,+\,\wp_{u,v}\,m^2 I\right)\,
\psi^\prime(t',\,{\vec x}\,'\,)\,=\,0\,.
$
It is a straight forward algebraic exercise to find that $S(P)$ must
simultaneously satisfy the following requirements
\begin{eqnarray}
S^{-1}(P)\,\gamma_{00}\,S(P)\,=\,\gamma_{00},
\quad
S^{-1}(P)\,\gamma_{0\jmath}\,S(P)\,=\,-\,
\gamma_{0\jmath}\quad,
\nonumber\\
S^{-1}(P)\,\gamma_{\jmath0}\,S(P)\,=\,-\,
\gamma_{\jmath0},
\quad
S^{-1}(P)\,\gamma_{\ell\jmath}\,S(P)\,=\,\gamma_{\ell\jmath}
\quad.
\end{eqnarray}
Referring to Eqs. (\ref{gm}), we now note that while $\gamma_{00}$
commutes  with $\gamma_{\ell \jmath}$ it
anticommutes with $\gamma_{0\jmath}$
\begin{equation}
\left[\gamma_{00},\,\gamma_{\ell\jmath}\right]\,=\,
\left[\gamma_{00},\,\gamma_{\jmath\ell}\right]\,=\,0,\quad
\left\{\gamma_{00},\,\gamma_{0\jmath}\right\}\,=\,
\left\{\gamma_{00},\,\gamma_{\jmath0}\right\}\,=\,0\quad.
\end{equation}
As a result, confining to the norm preserving transformations (and ignoring a
possible {\it global} phase factor
\footnotemark[2]
\footnotetext[2]{Such a global phase factor acquires crucial  importance for
constructing internally consistent theory Majorana-like $(j,0)\oplus(0,j)$
 fields.}),
we  identify $S(P)$ with
$\gamma_{00}$, yielding:
$
\psi'(t',\,{\vec x}\,') \,=\,\gamma_{00}\,\psi(t,\,\vec x)\quad
\Longleftrightarrow\quad
\psi'(t',\,{\vec x}\,') \,=\,\gamma_{00}\,\psi(t',\,-\,{\vec x}\,')
\,.
$
This prepares us to proceed to the field theoretic considerations.
The $(1,0)\oplus(0,1)$ matter field operator is defined by letting
$\sigma=0,\pm 1$ in the general expression (\ref{fo}).
The transformation properties of  the states
$
\vert\vec p,\,\sigma\rangle^u\,=\,a^\dagger_\sigma(\vec p\,)\,\vert\mbox{vac}
\rangle$ and
$
\vert\vec p,\,\sigma\rangle^v\,=\,b^\dagger_\sigma(\vec p\,)\,\vert\mbox{vac}
\rangle\,
$
are obtained from the condition
\begin{equation}
U(P)\,\Psi(t',\,\vec x\,'\,)\,U^{-1}(P)\,=\,
\gamma_{00}\,\Psi(t',\,-\,\vec x\,'\,)\label{uuin}
\quad,
\end{equation}
where $U(P)$ represents a unitary operator which governs the operation
of parity in the Fock space.
Using the definition of $\gamma_{00}$, Eqs. (\ref{gm}), and the
explicit expressions for the
$(1,0)\oplus(0,1)$ spinors $u_\sigma(\vec p\,)$ and $v_\sigma(\vec p\,)$
given by Eqs. (\ref{uv}), we find
\begin{equation}
\gamma_{00}\, u_\sigma(p')\,=\,+\,u_\sigma(p)\,,\quad
\gamma_{00}\, v_\sigma(p')\,=\,-\,v_\sigma(p)\quad,
\label{pt}
\end{equation}
with $p'$  the parity-transformed $p$  --- i.e.
for $p^\mu\,=\,(p^0,\,\vec p\,)$,
$p^{\prime\mu}\,=\,(p^0,\,-\vec p\,)$. The observation (\ref{pt})
when coupled with the requirement (\ref{uuin}) immediately yields the
transformation properties of the
particle-antiparticle
 creation operators:\\
$
U(P)\,a^\dagger_\sigma(\vec p\,)\, U^{-1}(P)
\,=\,+\,a^\dagger_\sigma(-\,\vec p\,)\,,\,\,
U(P)\,b^\dagger_\sigma(\vec p\,)\, U^{-1}(P)
\,=\,-\,b^\dagger_\sigma(-\,\vec p\,)\,.
$
Under the assumption that the vacuum is invariant under the parity
transformation, $U(P)\,\vert\mbox{vac} \rangle\,=\,\vert\mbox{vac} \rangle$, we
arrive at the result that the
\footnotemark[3]
\footnotetext[3]{The fuller justification for the terminology
``particle'' and ``antiparticle,''
apart from the convention  of what we call ``particle,''
will be realised when we consider the
the operation of $C$.}
 ``particles'' (described classically  by the
$u$-spinors) and ``antiparticles'' (described classically by the $v$-spinors)
have
opposite relative intrinsic parities:
$
U(P)\,\vert\vec p,\,\sigma\rangle^u\,=\,+\,\vert\,-\vec p, \sigma
\rangle^u \,,\quad
U(P)\,\vert\vec p,\,\sigma\rangle^v\,=\,-\,\vert\,-\vec p, \sigma
\rangle^v \,.
$
This is are precisely what we set out to prove. That is, the
$(1,0)\oplus(0,1)$ boson and anti-boson  have opposite relative intrinsic
parity. As a consequence the number of physical states, in comparison to the
description of a massive spin one particle by the Proca vector potential
$A^\mu(x)$, are {\it doubled}  from $(2j+1)=3$ to $2(2j+1)=6$ in agreement
\footnotemark[4]
\footnotetext[4]{While our agreement with BWW [1b] is complete, we
differ with Weinberg's [2, footnote 13]  claim that the $(j,0)\oplus(0,j)$
fields have {\it same} relative intrinsic parity for bosons. The disagreement
with Ref. \cite{SW} arises because it's author  did not realise that the
bosonic $v$-spinors are {\it not} solutions of the equation which he proposed.
The factor $\wp_{u,v}$ in Eq. (\ref{ea})  {\it is required for internal
consistency} in the theory.}
with  BWW's work [1b].

Next we consider the operation of C. The charge conjugation operation C
must be carried through with a little greater care for bosons than
for fermions within the $(j,0)\oplus(0,j)$ framework developed here because of
$\wp_{u,v}$ factor in the mass term. For the $(1,0)\oplus(0,1)$ case,
at the classical level we want
\begin{equation}
C:\quad
\left(\gamma_{\mu\nu}\,D^\mu_{(+)}\,D^\nu_{(+)}\,+\,m^2\right)\,u(x)\,=\,0
\,\,\,\longrightarrow\,\,\,
\left(\gamma_{\mu\nu}\,D^\mu_{(-)}\,D^\nu_{(-)}\,-\,m^2\right)\,v(x)\,=\,0
\quad,\label{c}
\end{equation}
where the local $U(1)$ gauge covariant derivatives are defined as:
$
D^\mu_{(+)}\,=\,\partial^\mu\,+\,i\,q\, A^\mu(x)\,,\quad
D^\mu_{(-)}\,=\,\partial^\mu\,-\,i\,q\, A^\mu(x)\,.
$
Again a straight forward algebraic exercise yields the
result that for (\ref{c}) to occur we must have:
$
\psi(t,\vec x) \,\,\,\rightarrow \,\,\,C\,\psi^\ast(t,\vec x)\,;
$
where $C$ satisfies,
$
C\,\gamma^\ast_{\mu\nu}\,C^{-1}\,=\,-\,\gamma_{\mu\nu}\,.
$
We find that
$
C\,=\,\eta_\sigma\,A\,\gamma_{00}\,,
$
with
\begin{equation}
A\,=\,\left(
\begin{array}{ccc}
0&{\,\,}&\Theta_{[1]}\\
\Theta_{[1]}&{\,\,}&0
\end{array}
\right)\,,\quad
\Theta_{[1]}\,=\,
\left(
\begin{array}{ccccc}
0&{\,\,}&0&{\,\,}&1\\
0&{\,\,}&-1&{\,\,}&0\\
1&{\,\,}&0&{\,\,}&0
\end{array}
\right)
\quad,
\end{equation}
and for convenience
we choose $\eta_{\pm 1}=+1$ and $\eta_{0}=-1$; and $\Theta_{[1]}$ is Wigner's
time reversal operator \cite{TO} for spin-$1$. The effect of
charge conjugation in the Fock space is now immediately obtained
by using the easily verifiable identities:
$
C\,u^\ast_{+1}(\vec p\,)\,=\,v_{-1}(\vec p\,)\,,\quad
C\,u^\ast_{0}(\vec p\,)\,=\,v_{0}(\vec p\,)\,,\quad
C\,u^\ast_{-1}(\vec p\,)\,=\,v_{+1}(\vec p\,)\,;
$
and the requirement
$
U(C)\,\Psi(x)\,U^{-1}(C)\,=\,\Psi^c(x)\,,
$
where
\begin{equation}
\Psi^c(x)\,=\,
\sum_{\sigma=0,\pm 1}
\int {d^3p\over (2\pi)^{3} } {1\over 2\,\omega_{\vec p}}
\Big[ S(C)\,u^\ast_\sigma(\vec p\,)\, a^\dagger_\sigma(\vec p\, )\,
e^{i p\cdot x}
+  S(C)\,v^\ast_\sigma(\vec p\,)
\,b_\sigma(\vec p\,) \,e^{-i p\cdot x} \Bigr ]
\quad.\label{pc}
\end{equation}
These considerations yield:
$
U(C)\,a^\dagger_\sigma(\vec p\,)\,U^{-1}(C)\,=\,b^\dagger_\sigma(\vec p\,)\,,
\quad
U(C)\,b^\dagger_\sigma(\vec p\,)\,U^{-1}(C)\,=\,a^\dagger_\sigma(\vec p\,)\,.
$
We thus see that the definition of charge conjugation operation
C as given by (\ref{c}) indeed yields the correct picture in the Fock space:
$
U(C)\,\vert\vec p,\sigma\rangle^u\,=\,\vert\vec p,\sigma\rangle^v
$ and
$
U(C)\,\vert\vec p,\sigma\rangle^v\,=\,\vert\vec p,\sigma\rangle^u\,.
$

Finally, following Nachtmann \cite{ON}, we define the operation of time
reversal as a {\it product}
\footnotemark[5]
\footnotetext[5]{The {\it order} of the operations in the product which follows
is not important because the two operations are found to anticommute, and
therefore the ambiguity  of ordering only involves  an overall global phase
factor.}
of an operation, $S^\prime(\Lambda_T)$,
$
\psi(t,\vec x\,)\,\,\longrightarrow \,\,S^\prime(T)\,\psi(-t,\vec x\,),
$
which preserves the
form of Eq. (\ref{ea}) under $x^\mu\rightarrow {[\Lambda_T]^\mu}_\nu\,
x^{\,\prime\,\nu}$, $\Lambda_T=diag\,(-1,\,1,\,1,\,1)$,  {\it and}, based on
St\"uckelberg-Feynman \cite{SF} arguments, the operation of charge conjugation.
So, classically, under $T$ we have:
$
\quad\psi(t,\vec x)\,\,\longrightarrow\,\,
\psi^\prime(t,\vec x) \,=\,S^\prime(\Lambda_T)\,S(C)\,
\psi^\ast(-t,\vec x\,)\,.
$
We find that $S(T) \,\equiv\,S^\prime(\Lambda_T)\,S(C)$ is given by:
$
S(T)\,=\,(A\,\, global\,\, phase\,\, factor)\,\times\,\gamma_{00}\,
 \eta_\sigma\,A\,\gamma_{00}\,.
$
Taking note of the fact that $A$ anticommutes with $\gamma_{00}$,
$\{A,\,\gamma_{00}\}=0$, and dropping the resultant
global phase factor, we obtain $S(T)=\eta_\sigma\,A$.
In the Fock space above considerations are implemented by finding the effect of
an {\it anti-unitary} operator on the creation and annihilation operators via:
$
\left[V(T)\,\Psi(t,\vec x\,)\,V^{-1}(T)\right]^\dagger
\,=\,\Psi^\prime(t,\vec x\,)\,,
$
where
\begin{equation}
\Psi^{\prime}(t,\vec x\,)
\,=\,
\sum_{\sigma=0,\pm 1}
\int {d^3p\over (2\pi)^{3} } {1\over 2\,\omega_{\vec p}}
\Big[S(T)\, u^\ast_\sigma(\vec p\,)\, a^\dagger
_\sigma(\vec p\, )\, e^{i p\cdot x'}
+ S(T)\,  v^\ast_
\sigma(\vec p\,) \,b_\sigma(\vec p\,) \,e^{-i p\cdot x'} \Bigr ]
\quad,\label{t}
\end{equation}
with $x^{\prime\,\mu}=(-t,\vec x\,)$. Exploiting  the \\
identities
$
S(T)\,u_\sigma^\ast(\vec p\,)\,=\,u_{-\sigma}(-\vec p\,)\,,\quad
S(T)\,v_\sigma^\ast(\vec p\,)\,=\,v_{-\sigma}(-\vec p\,)\,
$
we arrive at the result:
$
V(T)\,a^\dagger_\sigma(\vec p\,)\,V^{-1}(T)\,=\,
a^\dagger_{-\sigma}(-\vec p\,)\,,\quad
V(T)\,b^\dagger_\sigma(\vec p\,)\,V^{-1}(T)\,=\,
b^\dagger_{-\sigma}(-\vec p\,)\,.
$
Therefore if the vacuum is invariant under T, the physical states transform as
$V(T)\,\vert\vec p,\,\sigma\rangle \,=\, \vert-\vec p,\,-\sigma\rangle$ and
observables ${\cal O}\,\rightarrow {\cal O}^\prime\,=\, [V(T)\,{\cal O}\,
V^{-1}(T)]^\dagger$.

We thus see that the $(1,0)\oplus(0,1)$ field theory constructed above is
indeed of BWW-type. What is left, in view of the already cited
work of Lee and Wick \cite{LW},
is to explicitly show that the $(j,0)\oplus(0,j)$
field operator (\ref{fo}) for spin one
describes a {\it nonlocal} theory (in the sense
to become obvious below). A straight forward algebraic exercise yields
the result that for the $(1,0)\oplus(0,1)$ field operator associated
with massive particles
\begin{eqnarray}
\Big[&&\Psi_\alpha(t,\,\vec x\,)\,,\,\,\overline{\Psi}_\beta(t,\,
\vec x^{\,\prime}\,)
\Big]\,=\,\nonumber \\
&&\left( {1\over {2\,\pi}}\right)^6
\int  {{d^3\vec p}\over{2\,E(\vec p\,)} }
\sum_{\sigma=0,\pm 1} \bigg(
u_\sigma(\vec p\,)\,{\overline u}_\sigma(\vec p\,)
+
v_\sigma(-\,\vec p\,)\,{\overline v}_\sigma(-\,\vec p\,)
\bigg)_{\alpha\beta}\,
e^{i\vec p\cdot(\vec x-\vec x^\prime)}\quad.\label{nl}
\end{eqnarray}

The  nonlocality is now immediately inferred.
Using the explicit forms (\ref{uv}) of $u_{0,\pm 1}(\vec p\,)$
and $v_{0,\pm 1}(\vec p\,)$, we find
\begin{equation}
\sum_{\sigma=0,\pm 1} \bigg(
u_\sigma(\vec p\,)\,{\overline u}_\sigma(\vec p\,)
+
v_\sigma(-\,\vec p\,)\,{\overline v}_\sigma(-\,\vec p\,)
\bigg)\,=\,
\left(
\begin{array}{cc}
{\cal M} & 0\\
0 & -{\cal M}
\end{array}\right)
\quad,
\end{equation}
with
\begin{equation}
{\cal M}\,=\,
\left(
\begin{array}{ccccc}
E^2\,+\,p_z^2 &{\;\;}& \sqrt{2} \,p_-\, p_z&{\;\;} & p_-^2 \\
\sqrt{2}\, p_+\, p_z &{\;\;}& E^2\,+\,p_-\,p_+\, -\,p_z^2& {\;\;}&
-\,\sqrt{2}\, p_-\, p_z \\
p_+^2 &{\;\;}& -\,\sqrt{2} \,p_+\, p_z&{\;\;} & E^2\,+\,p_z^2
\end{array}
\right)\label{m}\quad.
\end{equation}
In Eq. (\ref{m}) $p_\pm=p_x\pm i p_y$.
Consequently:
$\left[\Psi_\alpha(t,\,\vec x\,)\,,\,\,\overline{\Psi}_\beta(t,\,\vec
x^\prime\,) \right]\,\ne\, (const.)\times\delta^3(\vec x-\vec x^\prime\,)$.
For comparison we note that a
 similar calculation for the spin half Dirac case yields
$\left\{\Psi_\alpha(t,\,\vec x\,)\,,\,\,\overline{\Psi}_\beta(t,\,\vec
x^\prime\,)
\right\}\,=\,(const.)\times
 \gamma^0_{\alpha\beta}\,\delta^3(\vec x-\vec x^{\,\prime}\,)\,$. The crucial
property of the $(1/2,0)\oplus(0,1/2)$ representation space which enters
in obtaining this result is :
$
\sum_{\sigma=\pm {1\over 2}}\Big(
u_\sigma(\vec p\,)\,{\overline u}_\sigma(\vec p\,)
\,+\,
v_\sigma(-\,\vec p\,)\,{\overline v}_\sigma(-\,\vec p\,)
\Big)_{\alpha\beta} \sim \left(\gamma^0\right)_{\alpha\beta}\,\,p_0
\,.
$
No corresponding {\it overall} factor of $E$ appears in
$\cal M$. If it did, the $E(\vec p\,)^{-1}$ factor in the
invariant element of phase space could be cancelled leading to a
$\delta^3(\vec x-\vec x^\prime\,)$ and its derivatives
(Schwinger terms \cite{JS}) on the {\it rhs} of Eq. (\ref{nl})
thus restoring  the locality.
For an alternate derivation of the nonlocality the reader may
wish to refer to Ref. [3b]. For the sake of completeness we note that
the momentum conjugate to the field operator $\Psi(x)$ for spin one
is given by $\Pi_0(x)=\overline{\Psi}(x)\,\gamma_{\mu 0}\,\partial^\mu$, and
$
\left[\Psi_\alpha(t,\,\vec x\,)\,,\,\,(\Pi_0)_\beta (t,\,\vec x^{\,\prime})
\right]\,=\,
-\,\delta^3(\vec x\,-\,\vec x^{\,\prime})\,
\partial^\mu\Psi_\alpha(t,\,\vec x\,)\,
\overline{\Psi}_\xi(t,\,\vec x\,)\,(\gamma_{\mu 0})_{\xi\beta}
\,.$ The physical interpretation of this last result requires further study.

To summarise we note the $(j,0)\oplus(0,j)$ representation space associated
with massive particles is a concrete realisation of a quantum field theory,
envisaged many years ago by Bargmann, Wightman and Wigner, in which bosons and
antibosons have opposite relative intrinsic parities. It is  our hope that our
detailed analysis of the $(j,0)\oplus(0,j)$ representation space would
supplement the canonical Bargmann-Wigner/Rarita-Schwinger \cite{BWRS} formalism
(where a boson and anti-boson have same intrinsic parity) and open new
experimentally observable possibilities for the Poincar\'e covariant aspects of
hadrons and their propagation in nuclei.

\acknowledgements
It is with pleasure that we  thank an  anonymous referee for an unusually
detailed report which contained  many helpful questions and comments. In
particular, the referee helped us note  that for the  $C$ and $P$ operators
defined in the manuscript $CP\,=\,\bbox - PC$. As a consequence, the particles
and antiparticles in the $(1,0)\oplus(0,1)$ representation space do indeed have
opposite intrinsic parity and that our conclusion that we have constructed a
Bargmann-Wightman-Wigner type quantum field theory is independent of any
choice/conventions of phase factors.

\end{document}